\documentclass[12pt]{iopart}
\usepackage{iopams}
\usepackage{graphicx}
\usepackage{epsf}
\usepackage{bm}

\usepackage{dcolumn}

\newcolumntype{.}{D{x}{}{-1}}

\begin{document}

\newcommand{\half}{\frac12}
\newcommand{\vare}{\varepsilon}
\newcommand{\pr}{^{\prime}}
\newcommand{\ppr}{^{\prime\prime}}
\newcommand{\pp}{{p^{\prime}}}
\newcommand{\hp}{\hat{\bfp}}
\newcommand{\hpp}{\hat{\bfpp}}
\newcommand{\hx}{\hat{\bfx}}
\newcommand{\hq}{\hat{\bfq}}
\newcommand{\hz}{\hat{\bfz}}
\newcommand{\hr}{\hat{\bfr}}
\newcommand{\hn}{\hat{\bfn}}
\newcommand{\rx}{{\rm x}}
\newcommand{\rp}{{\rm p}}
\newcommand{\rmq}{{\rm q}}
\newcommand{\rpp}{{{\rm p}^{\prime}}}
\newcommand{\rk}{{\rm k}}
\newcommand{\bfe}{{\bm e}}
\newcommand{\bfp}{{\bm p}}
\newcommand{\bfpp}{{\bm p}^{\prime}}
\newcommand{\bfq}{{\bm q}}
\newcommand{\bfx}{{\bm x}}
\newcommand{\bfk}{{\bm k}}
\newcommand{\bfy}{{\bm y}}
\newcommand{\bfz}{{\bm z}}
\newcommand{\bfr}{{\bm r}}
\newcommand{\bfn}{{\bm n}}
\newcommand{\intzo}{\int_0^1}
\newcommand{\intinf}{\int^{\infty}_{-\infty}}
\newcommand{\ka}{\kappa_a}
\newcommand{\kb}{\kappa_b}
\newcommand{\ThreeJ}[6]{
        \left(
        \begin{array}{ccc}
        #1  & #2  & #3 \\
        #4  & #5  & #6 \\
        \end{array}
        \right)
        }
\newcommand{\SixJ}[6]{
        \left\{
        \begin{array}{ccc}
        #1  & #2  & #3 \\
        #4  & #5  & #6 \\
        \end{array}
        \right\}
        }
\newcommand{\NineJ}[9]{
        \left\{
        \begin{array}{ccc}
        #1  & #2  & #3 \\
        #4  & #5  & #6 \\
        #7  & #8  & #9 \\
        \end{array}
        \right\}
        }
\newcommand{\Dmatrix}[4]{
        \left(
        \begin{array}{cc}
        #1  & #2   \\
        #3  & #4   \\
        \end{array}
        \right)
        }
\newcommand{\cross}[1]{#1\!\!\!/}
\newcommand{\eps}{\epsilon}
\newcommand{\beq}{\begin{equation}}
\newcommand{\eeq}{\end{equation}}
\newcommand{\beqn}{\begin{eqnarray}}
\newcommand{\eeqn}{\end{eqnarray}}
\newcommand{\lbr}{\langle}
\newcommand{\rbr}{\rangle}
\newcommand{\Za}{Z\alpha}

\title[Nuclear-size self-energy and vacuum-polarization corrections]{Nuclear-size self-energy and vacuum-polarization corrections to the bound-electron $\bm{g}$ factor}

\author{V.~A. Yerokhin}
\address{Max~Planck~Institute for Nuclear Physics, Saupfercheckweg~1, D~69117 Heidelberg,
Germany} \address{ExtreMe Matter Institute EMMI, GSI Helmholtzzentrum f\"ur
Schwerionenforschung, D-64291 Darmstadt, Germany} \address{Center for Advanced Studies,
St.~Petersburg State Polytechnical University, Polytekhnicheskaya 29,
        St.~Petersburg 195251, Russia}

\author{C.~H. Keitel}
\address{Max~Planck~Institute for Nuclear Physics, Saupfercheckweg~1, D~69117 Heidelberg, Germany}

\author{Z. Harman}
\address{Max~Planck~Institute for Nuclear Physics, Saupfercheckweg~1, D~69117 Heidelberg, Germany}

\begin{abstract}

The finite nuclear-size effect on the leading bound-electron $g$ factor and the one-loop QED
corrections to the bound-electron $g$ factor is investigated for the ground state of hydrogen-like ions.
The calculation is performed to all orders
in the nuclear binding strength parameter $\Za$ (where $Z$ is the nuclear charge and $\alpha$
is the fine structure constant) and for the Fermi model of the nuclear charge distribution.
In the result, theoretical predictions for the isotope shift of the $1s$
bound-electron $g$ factor are obtained, which can be used for the determination of the difference of nuclear charge radii
from experimental values of the bound-electron $g$ factors for different isotopes.

\end{abstract}


\maketitle

Significant progress has been achieved during the last two decades in the experimental
determination of the bound-electron $g$ factor in hydrogen-like (and lithium-like) ions
\cite{haeffner:00:prl,verdu:04,sturm:11,sturm:13:Si,wagner:13}. The current experimental precision
is on the level of few parts in $10^{-11}$ and is likely to be improved further in the future.
Comparison between the experimental and theoretical results constituted a highly
sensitive test of bound-state QED theory \cite{yerokhin:02:prl,pachucki:04:prl,pachucki:05:gfact}
and led to an accurate determination of the electron mass \cite{beier:02:prl,mohr:13:codata}.
In future, such experiments can also provide us with a new method
of determination of other important parameters, in particular,
the fine-structure constant \cite{shabaev:06:prl} and nuclear magnetic moments
\cite{yerokhin:11:prl}.

In the present work, we investigate one of the possibilities opened by the high-precision $g$
factor experiments, namely, a possibility to determine the nuclear charge radius or the
difference of the nuclear charge radii of two isotopes. A proof-of-the-principle determination of
the charge radius
of $^{28}$Si has already been reported in the recent $g$-factor measurement \cite{sturm:13:Si}.
The nuclear charge distribution effect will play a much more significant role
when the planned extension of the $g$-factor measurements to higher-$Z$ systems \cite{blaum:priv}
takes place.

At the present level of theory, the direct determination of the nuclear charge radius
is restricted by the theoretical uncertainty due to the two-loop QED effects
\cite{pachucki:04:prl,pachucki:05:gfact,yerokhin:13:twoloopg}. In order to avoid this
restriction, it might be advantageous to
study the isotopic difference of the bound-electron $g$ factor values.
Theoretical description of the isotope shift of the $g$ factors is much
simpler than that of the $g$ factor itself,
as many corrections (in particular, the dominant part of
the two-loop QED effects) do not depend on nuclear properties and cancel in the difference.
The first experimental determination of the isotopic shift of the bound-electron $g$ factor is
currently underway for a calcium ion \cite{blaum:priv}.

The goal of the present work is to
perform a detailed investigation of the finite nuclear-size effect on the
leading bound-electron $g$ factor and on the one-loop QED corrections to the $g$ factor.
The results obtained, combined with the previously reported data on the nuclear recoil
correction, allow one to deduce accurate values for the nuclear-dependent part of the $1s$
bound-electron $g$ factor and, therefore, the isotope shift of the $g$ factor.

The remaining paper is organized as follows. In the next section, we discuss the nuclear-size
correction to the leading-order bound-electron $g$ factor. In Sec.~\ref{sec:2}, we calculate the
nuclear-size effect on the self-energy and vacuum-polarization corrections to the $g$ factor.
Numerical results and experimental consequences are summarized and discussed in Sec.~\ref{sec:3}.
The relativistic units ($\hbar=c=1$) are used throughout the paper.
\section{Nuclear-size correction to the leading-order $\bm{g}$ factor}
\label{sec:1}

We start with the nuclear-size correction to the relativistic (Breit) value of the bound-electron $g$ factor, defined by
the difference
\begin{equation} \label{eq:01}
\delta g_{\rm N} = g_{\rm ext}^{(0)} - g_{\rm pnt}^{(0)}\,,
\end{equation}
where $g_{\rm ext}^{(0)}$ and $g_{\rm pnt}^{(0)}$ are the leading-order bound-electron $g$ factor
values evaluated with the extended and the point nuclear models, respectively. For the point
nucleus, the well-known analytical result for the $1s$ state reads
\begin{equation} \label{eq:02}
g_{\rm pnt}^{(0)} = \frac23\left[ 1 + 2\,\sqrt{1-(\Za)^2}\right]\,,
\end{equation}
whereas for the extended nucleus the $g$ factor value is given (for the $1s$ state) by the integral
\begin{equation} \label{eq:03}
g_{\rm ext}^{(0)} = -\frac83\, \int_0^{\infty}dr\,r^3\,g_a(r)\,f_a(r)\,,
\end{equation}
where $g_a$ and $f_a$ are the upper and lower radial components of the (extended-nucleus)
reference-state wave function.

The nuclear-size correction $\delta g_{\rm N}$ can be readily evaluated numerically \cite{persson:97:g,beier:00:rep}.
For light ions, it can be also obtained analytically by using the expansion in the nuclear binding strength
parameter $\Za$ \cite{karshenboim:00:pla,glazov:01:pla}. In Ref.~\cite{karshenboim:05}, a simple
approximate relation was established between the nuclear-size corrections to the $g$ factor and to
the binding energy. For the $1s$ state, it reads
\begin{equation} \label{eq:03a}
\delta g_{\rm N} = \frac43\,(2\gamma+1)\,\frac{\delta E_{\rm N}}{m}\,,
\end{equation}
where $\delta E_{\rm N}$ is the leading-order nuclear-size correction to the Dirac energy and $\gamma =
\sqrt{1-(\Za)^2}$. The relation (\ref{eq:03a}) goes beyond the $\Za$ expansion and holds with good
accuracy in the whole region of the nuclear charge numbers $Z$.

In Table~\ref{tab:fns}, we present our
numerical results for the nuclear-size correction $\delta g_{\rm N}$ to the $1s$ bound-electron $g$
factor. The results are parameterized in terms of the dimensionless function $G_{\rm N}(Z,R)$
defined as
\begin{equation} \label{eq:04}
\delta g_{\rm N} = \frac83\, (\Za)^4\, \left( \frac{R}{\lambda_C} \right)^{2\gamma}\,G_{\rm N}\,,
\end{equation}
where $\lambda_C = \hbar/mc$, and $R \equiv \lbr r^{2}\rbr^{1/2}$ is the root-mean-square
(rms) radius of the nuclear-charge distribution. The prefactor before $G_{\rm N}$ in
Eq.~(\ref{eq:04})
is consistent with the leading term of the $\Za$ expansion of $\delta g_{\rm N}$ \cite{karshenboim:00:pla},
$(8/3)(\Za)^4(R/\lambda_C)^2$, so that $G_{\rm N}$ is unity in the nonrelativistic limit.
The exponent of $R$ in Eq.~(\ref{eq:04}) follows from Eq.~(\ref{eq:03a}) and the relativistic
result for the nuclear-size correction to the energy obtained in Ref.~\cite{shabaev:93:fns}.

The numerical values of the function $G_N$ are presented in the third column of
Table~\ref{tab:fns}. The results are obtained with the standard two-parameter Fermi model for the
nuclear charge distribution (with the standard choice of the thickness parameter $t = 2.3$~fm).
Nuclear rms radii used in the calculation are listed in the second column of the table.
They were taken from Ref.~\cite{angeli:04}, with the only exception of $Z=92$ for which we used
the value from Ref.~\cite{kozhedub:08}.

We observe that the function $G_{\rm N}$ stays remarkably close to unity in the whole range of $Z$.
It might be noted that such smooth behavior of $G_{\rm N}$ is a consequence of the correct relativistic
exponent of $R$ in Eq.~(\ref{eq:04}). If we used $R^2$
instead of $R^{2\gamma}$ in Eq.~(\ref{eq:04}), we would get a much more rapidly varying function.
Namely, $(R/\lambda_C)^{2\gamma-2} \approx 14$ for $Z = 100$, so $G_{\rm N}(Z = 100)$ would have been 14 times larger
within the $R^2$ parametrization.

Having accurate numerical results for $\delta g_N$, it might be interesting to
check how well the approximate relation (\ref{eq:03a}) holds. Our calculation shows that this relation
is accurate to about 1\% in the high-$Z$ region and better than that in the low-$Z$ region. Namely,
with the numerical results for $\delta E_{\rm N}$ from Ref.~\cite{yerokhin:11:fns},
Eq.~(\ref{eq:03a}) yields $G_{\rm N}(Z = 100) = 1.125$, $G_{\rm N}(Z = 82) = 1.213$,
and $G_{\rm N}(Z = 40) = 1.119$, which can be compared with the exact numerical results in the third column of
Table~\ref{tab:fns}.

In the fourth column of Table~\ref{tab:fns}, we present differences of the
results for $G_N$ obtained with the Fermi and the homogeneously charged nuclear models. This
difference can be considered as an estimate of the model dependence of the calculational results for
the finite nuclear-size correction. We observe that the model dependence of the results is
rather weak, ranging from  $0.03\%$ in the low-$Z$ region to $0.2\%$ in the high-$Z$ region.

The leading dependence of the nuclear-size correction on the nuclear radius is factorized out by the
prefactor $R^{2\gamma}$ in Eq.~(\ref{eq:04}). Still, the function $G_N$ depends on $R$, albeit
weakly. In order to estimate its $R$ dependence, in the last column of Table~\ref{tab:fns} we list
the results for the derivative $G^{\prime}_{\rm N}(R) = dG_{\rm N}(R)/dR$. We observe that the derivative
$G^{\prime}(R)$ is small and scales almost linearly with $Z$.

Numerical data for $G_{\rm N}$ and $G^{\prime}_{\rm N}$ listed in Table~\ref{tab:fns}
allow one to obtain accurate results for the nuclear-size correction
to the isotope shift of the $g$ factor.
E.g., the difference of the nuclear-size corrections for $^{208}$Pb (with $R = 5.5010$~fm) and
$^{204}$Pb (with $R = 5.4794$~fm) calculated by using the values of $G_{\rm N}$ and $G^{\prime}_{\rm N}$
listed in Table~\ref{tab:fns} is $2.8078 \times 10^{-6}$, which agrees to all digits with the
direct numerical evaluation. The corresponding result calculated without $G^{\prime}_{\rm N}$
($2.846 \times 10^{-6}$) is much less accurate.

Numerical results for $G^{\prime}_{\rm N}$ can also be used for estimating the nuclear deformation effects
on the bound-electron $g$ factor. It was demonstrated \cite{zatorski:12} that the leading quadrupole and
hexadecapole nuclear deformation effects to the $g$ factor can be parameterized in terms of shifts of the rms radius.
In particular, using Eq.~(11) of Ref.~\cite{zatorski:12} one can easily determine the correction to
the rms radius due to the quadrupole $\beta_2$ and hexadecapole $\beta_4$ nuclear deformation
parameters and then, using the numerical values for $G^{\prime}_{\rm N}$ from Table~\ref{tab:fns},
obtain the corresponding corrections to the $g$ factor.

\begin{table}
\caption{Leading-order nuclear-size correction to the $1s$ bound-electron $g$ factor. Notations are
as follows:
$R$ is the nuclear-charge root-mean-square radius (in fermi),
$G_{\rm N}$ is the nuclear-size correction for the Fermi model of the nuclear charge distribution,
$\delta G_{\rm N}$ is the difference between the results obtained with the Fermi and the
homogeneously charged sphere models, and $G^{\prime}_{\rm N}$ is the derivative of $G_{\rm N}$ with
respect of $R$ (with $R$ being expressed in fermi units).
 \label{tab:fns} }
\begin{center}
\begin{tabular}{lc.....}
\hline\\[-0.5cm]
                $Z$ & R [fm]
                & \multicolumn{1}{c}{$G_{\rm N}$}
                & \multicolumn{1}{c}{$\delta G_{\rm N}$}
                & \multicolumn{1}{c}{$G^{\prime}_{\rm N}$ [1/fm]}
 \\
\hline\\[-9pt]
  4 & 2.5180 &  1.0x02  &          &  \\
  6 & 2.4703 &  1.0x05  &          &  \\
  8 & 2.7013 &  1.0x09  &          &  \\
 10 & 3.0053 &  1.0x13  &          &  \\
 12 & 3.0568 &  1.0x183 & -0.0x003 &  \\
 14 & 3.1223 &  1.0x237 & -0.0x003 &  \\
 16 & 3.2608 &  1.0x295 & -0.0x004 &  \\
 18 & 3.4269 &  1.0x356 & -0.0x004 & -0.0x005 \\
 20 & 3.4764 &  1.0x421 & -0.0x005 & -0.0x006 \\
 24 & 3.6424 &  1.0x559 & -0.0x006 & -0.0x007 \\
 30 & 3.9286 &  1.0x781 & -0.0x008 & -0.0x010 \\
 32 & 4.0744 &  1.0x857 & -0.0x008 & -0.0x011 \\
 36 & 4.1882 &  1.1x013 & -0.0x010 & -0.0x013 \\
 40 & 4.2696 &  1.1x170 & -0.0x011 & -0.0x014 \\
 44 & 4.4818 &  1.1x324 & -0.0x012 & -0.0x017 \\
 48 & 4.6137 &  1.1x473 & -0.0x014 & -0.0x019 \\
 50 & 4.6543 &  1.1x545 & -0.0x015 & -0.0x020 \\
 54 & 4.7866 &  1.1x681 & -0.0x016 & -0.0x023 \\
 58 & 4.8770 &  1.1x806 & -0.0x018 & -0.0x025 \\
 60 & 4.9118 &  1.1x862 & -0.0x018 & -0.0x027 \\
 64 & 5.1617 &  1.1x955 & -0.0x019 & -0.0x031 \\
 68 & 5.2505 &  1.2x030 & -0.0x020 & -0.0x034 \\
 70 & 5.3115 &  1.2x057 & -0.0x020 & -0.0x036 \\
 74 & 5.3670 &  1.2x090 & -0.0x022 & -0.0x039 \\
 78 & 5.4278 &  1.2x087 & -0.0x023 & -0.0x043 \\
 80 & 5.4633 &  1.2x070 & -0.0x023 & -0.0x045 \\
 82 & 5.5010 &  1.2x042 & -0.0x024 & -0.0x048 \\
 83 & 5.5211 &  1.2x023 & -0.0x024 & -0.0x049 \\
 88 & 5.6841 &  1.1x877 & -0.0x024 & -0.0x055 \\
 90 & 5.7100 &  1.1x794 & -0.0x024 & -0.0x057 \\
 92 & 5.8569 &  1.1x689 & -0.0x023 & -0.0x060 \\
100 & 5.8570 &  1.1x115 & -0.0x023 & -0.0x068 \\
\hline
\end{tabular}
\end{center}
\end{table}

\section{Nuclear-size QED corrections}
\label{sec:2}

\begin{figure}
\centerline{
\resizebox{\textwidth}{!}{%
  \includegraphics{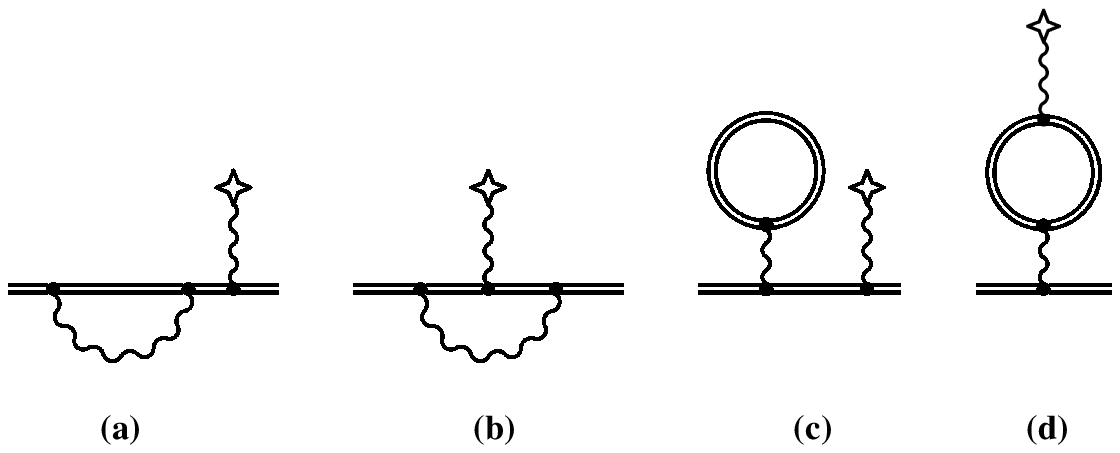}
}}
 \caption{Feynman diagrams representing QED corrections to the bound-electron $g$ factor. The
 self-energy is represented by graphs (a) and (b), the electric-loop vacuum-polarization by graph (c),
 the magnetic-loop
 vacuum-polarization by graph (d). Double
lines denote an electron propagating in the binding nuclear field, wave lines denote virtual
photons, and the wave line terminated by a cross denotes interaction with an external magnetic
field.
 \label{fig:qed_gfact}}
\end{figure}

The nuclear-size QED correction to the bound-electron $g$ factor can be conveniently parameterized
in terms of the dimensionless function $G_{\rm NQED}(Z,R)$,
\begin{equation} \label{eq:1}
\delta g_{\rm NQED} = \delta g_{\rm N}\, \frac{\alpha}{\pi}\, G_{\rm NQED}\,,
\end{equation}
where $\delta g_{\rm N}$ is the leading-order nuclear-size correction given by Eq.~(\ref{eq:04}).
Such parametrization of the nuclear-size QED effect is similar
to the one used for the Lamb shift \cite{milstein:04,yerokhin:11:fns}.
The function $G_{\rm NQED}(Z)$ will be divided into several parts,
\begin{equation}
G_{\rm NQED} = G_{\rm NSE} + G_{\rm NVP,el} + G_{\rm NVP, ml}\,,
\end{equation}
where $G_{\rm SE}$ is the nuclear-size correction to the self-energy and
$G_{\rm NVP,el}$ and $G_{\rm NVP,ml}$ are
the nuclear-size corrections to the electric-loop and magnetic-loop vacuum-polarization,
correspondingly. The self-energy correction
to the bound-electron $g$ factor is represented graphically on Fig.~\ref{fig:qed_gfact} (a) and (b),
whereas the electric-loop and magnetic-loop
vacuum-polarization corrections are represented by Fig.~\ref{fig:qed_gfact} (c) and (d),
correspondingly.

The nuclear-size effect on the QED corrections to the bound-electron $g$ factor was
taken into account previously in several studies. Namely, it was included into the
self-energy and vacuum-polarization calculations of Refs.~\cite{persson:97:g,beier:00:rep}
and into the self-energy calculation of Ref.~\cite{yerokhin:04}.
In Ref.~\cite{karshenboim:05}, an approximate relation was obtained between the
nuclear-size corrections to the $g$ factor and to the binding energy.
According to that work, the relative values of the nuclear-size vacuum-polarization corrections
to the $1s$ $g$-factor and to the $1s$ binding energy are equal (within the leading logarithmic approximation),
\begin{equation} \label{eq:33a}
\frac{\delta g_{\rm NVP}}{\delta g_{\rm N}} \approx \frac{\delta E_{\rm NVP}}{\delta E_{\rm N}}\,,
\end{equation}
where $\delta E_{\rm NVP}$ is the nuclear-size vacuum-polarization correction to the energy.
In the present work, we calculate the
nuclear-size QED correction with a realistic Fermi model of the nuclear charge distribution
and achieve higher numerical accuracy than in previous studies.

The nuclear-size correction to the self-energy is calculated as the difference of the self-energy
corrections calculated with the extended and the point nuclear models. The general scheme of calculation of
the one-loop self-energy correction to the bound-electron $g$ factor was developed and described in detail
in the previous studies involving one of us \cite{yerokhin:02:prl,yerokhin:04}.
For the evaluation of the nuclear-size correction to the self-energy reported in the present work, we needed
to extended this scheme for the case of
the general binding potential. To this end, we employed the numerical approach for the evaluation of
the Dirac Green function for the arbitrary spherically symmetric potential (behaving as $\sim 1/r$
for $r\to \infty$) described in Ref.~\cite{yerokhin:11:fns}.

\subsection{Electric-loop vacuum-polarization}

The electric-loop vacuum-polarization correction to the bound-electron $g$ factor is represented by Fig.~\ref{fig:qed_gfact}(c)
and given by the following expression
\begin{equation}\label{eq:20}
\Delta g_{\rm VP, el} = 2\,\lbr a| \left[ V_{\rm Uehl} + V_{\rm WK}\right] | \delta_g a\rbr\,,
\end{equation}
where $|a\rbr$ is the reference-state wave function with a fixed momentum projection $\mu = 1/2$,
$|\delta_{g}
a\rbr$ is first-order perturbation of the reference-state wave function by
the effective $g$-factor potential $V_g = 2m\,[\bm{r}\times\bm{\alpha}]_z $,
\begin{equation}
|\delta_{g} a\rbr = \sum_{n\neq a} \frac{|n\rbr\lbr n|V_{g}|a\rbr}{\vare_a-\vare_n}\,,
\end{equation}
and $V_{\rm Uehl}$ and $V_{\rm WK}$ are the one-loop Uehling and Wichmann-Kroll potentials, respectively.
The Uehling potential is given by the well-known expression
\begin {eqnarray} \label{uehlexpr}
V_{\rm Uehl}(r)&=&-\Za \,\frac{2\alpha}{3\pi} \int\limits_0^\infty dr'\; 4\pi r'\rho (r') \nonumber
\\ &&\times \int\limits_1^\infty dt \; \left(1 +\frac{1}{2t^2}\right)
\frac{\sqrt{t^2-1}}{t^{2}}
\frac{e^{-2m|r-r'|t}-e^{-2m(r+r')t}}
{4mrt} \,,\nonumber \\
\end{eqnarray}
where
$Z\rho(r)$ is the density of the nuclear charge distribution ($\int \rho(r) d{\bf r}=1$).
The Wichmann-Kroll potential is given by \cite{soff:88:vp,manakov:89:zhetp}
\begin{eqnarray} \label{vp3}
V_{\rm WK}(r)  &= &\,\frac{2\alpha}{\pi}\,{\rm
  Re}\,\sum_{\kappa}|\kappa|\, \int_0^{\infty}d\omega\,
 \nonumber \\ && \times
   \int_r^{\infty}dr^{\prime} r^{\prime}\, \left(1-\frac{r^{\prime}}{r}
   \right)\,  {\rm Tr}\, G_{\kappa}^{(2+)}(i\omega,r^{\prime},r^{\prime})\,,
\end{eqnarray}
where $G_{\kappa}^{(2+)}$ is the Dirac-Coulomb Green function containing two or more interactions with the
binding nuclear field and "Tr" denotes the trace of the matrix.

The nuclear-size effect on the electric-loop vacuum-polarization correction was calculated as the
difference of the vacuum-polarization corrections given by Eq.~(\ref{eq:20}) evaluated with the
extended and the point nuclear models. Numerical calculation was carried out similarly to that for the
nuclear-size vacuum-polarization correction to the Lamb shift in Ref.~\cite{yerokhin:11:fns}.

\subsection{Magnetic-loop vacuum-polarization}

The magnetic-loop vacuum-polarization correction to the bound-electron $g$ factor is represented by Fig.~\ref{fig:qed_gfact}(d)
and given by the following expression
\begin{equation}\label{eq:30}
\Delta g_{\rm VP, ml} = \lbr a| V_{\rm VP,ml} | a\rbr\,,
\end{equation}
where $V_{\rm VP,ml}$ is the magnetic-loop vacuum-polarization potential \cite{artemyev:01},
\begin{eqnarray}\label{eq:32}
V_{\rm VP,ml}(\bfx) &= &\ \frac{i\alpha}{2\pi}\int_{-\infty}^{\infty}d\omega\,
 \int d\bfy\, d\bfz\, \frac{\balpha}{|\bfx-\bfy|}
   \nonumber \\ && \times
  {\rm Tr}\biggl[ \balpha\, G(\omega,\bfy,\bfz)\, V_g(\bfz)\, G(\omega,\bfz,\bfy)
   \nonumber \\ &&
  - \balpha\, G^{(0)}(\omega,\bfy,\bfz)\,  V_g(\bfz)\, G^{(0)}(\omega,\bfz,\bfy) \biggr]\,.
\end{eqnarray}
Here, $G(\omega,\bfx_1,\bfx_2)$ is the Dirac-Coulomb Green function and
$G^{(0)}(\omega,\bfx_1,\bfx_2)$ is the free Dirac Green function. The scalar product between the vectors of the $\balpha$ matrices is implicit in
Eq.~(\ref{eq:32}). It is assumed that the expectation value of the potential $V_{\rm VP,ml}$
is calculated with the reference-state wave functions with the momentum projection $\mu_a = 1/2$.
Note that magnetic-loop vacuum-polarization potential contains only the Wichmann-Kroll
contribution, as the Uehling part vanishes due to symmetry reasons.

After integrating over the angular variables and rotating the contour of the $\omega$ integration,
the magnetic-loop vacuum-polarization  correction to the $g$ factor can be expressed as (for an $ns$ reference state)
\begin{eqnarray}\label{eq:33}
\Delta g_{\rm VP,ml} & = & \frac{\alpha}{\pi}\int_0^{\infty}d\omega\,
dx\,dy\,dz\,z^3\min(x^3,y^3)\,
 \nonumber \\ && \times
g_a(x)f_a(x)
\sum_{\kappa_1\kappa_2} \frac49\, (\kappa_1+\kappa_2)^2\, \left[C_1(-\kappa_1,\kappa_2)\right]^2\,
 \nonumber \\ && \times
\biggl[G_{\kappa_1}^{11}G_{\kappa_2}^{22} + G_{\kappa_1}^{22}G_{\kappa_2}^{11} + G_{\kappa_1}^{12}G_{\kappa_2}^{21}
+ G_{\kappa_1}^{21}G_{\kappa_2}^{12}
 \nonumber \\ &&
- \ldots G \to G^{(0)}\ldots \biggr]\,,
\end{eqnarray}
where $G_{\kappa}^{ij} \equiv G_{\kappa}^{ij}(i\omega,y,z)$ is the radial component of the
Dirac-Coulomb Green function and the second term in the brackets is obtained from the first one by
substituting the Dirac-Coulomb Green function with the free Dirac Green function. The angular
coefficient $C_J(\kappa_a,\kappa_b)$ is given by
\begin{eqnarray} \label{CJ}
C_J(\kappa_a ,\kappa_b) & =& (-1)^{j_a+1/2} \,\sqrt{(2j_a+1)(2j_b+1)}\,
  \nonumber \\ && \times
 \ThreeJ{j_a}{J}{j_b}{\half}{0}{-\half}
 \frac{1 + (-1)^{l_a+l_b+J}}{2} \, ,
\end{eqnarray}
where $j = |\kappa|-1/2$ and $l = |\kappa+1/2|-1/2$.

Direct numerical calculations of Eqs.~(\ref{eq:30})-(\ref{eq:33}) to all orders in $Z\alpha$ have been performed in
Refs.~\cite{persson:97:g,beier:00:rep}. The calculation reported in these studies
was seriously complicated by slow convergence of the partial-wave
expansion, especially in the low-$Z$ region, which restricted the final numerical accuracy of the obtained results.
More recently, it was demonstrated \cite{lee:05} that
the dominant part of this correction (induced by the light-by-light scattering diagram)
can be obtained in a closed form, without any partial-wave expansion. The corresponding expression
reads
\begin{eqnarray}\label{eq:4}
\Delta g_{\rm VP,ml}({\rm appr}) & =&
-\frac{32}{3}\frac{\alpha (\Za)^2}{\pi}\,\int_0^{\infty} dq\,
  F(q)\,
 \nonumber \\&&  \times
\int_0^{\infty}dr \left( \frac{\sin qr}{qr}-\cos qr\right)\, r\,g_a(r)f_a(r)
\,,
\end{eqnarray}
where the function $F(q)$ was calculated numerically and tabulated in Ref.~\cite{lee:05}.

In order to calculate the nuclear-size correction to the magnetic-loop vacuum-polarization correction, we need to
calculate Eq.~(\ref{eq:30}) with an extended and point nuclear models and take the difference
of the two results. In order to simplify the numerical evaluation, we divide the nuclear-size
correction into two parts, the one induced by the nuclear-size effect on the reference-state wave
function and the one induced by the nuclear-size effect on the vacuum-polarization potential.
Symbolically, we can write this as
\begin{eqnarray}\label{eq:40}
\Delta g_{\rm NVP,ml} & = &\int dx \biggl[ g_a^{\rm ext}(x)f^{\rm ext}_a(x)-g_a^{\rm pnt}(x)f^{\rm
pnt}_a(x)\biggl]\,V_{\rm VP,ml}^{\rm pnt}(x)
\nonumber \\ &&
+ \int dx\,  g_a^{\rm ext}(x)f^{\rm ext}_a(x)\,\biggl[V_{\rm VP,ml}^{\rm ext}(x)-V_{\rm VP,ml}^{\rm pnt}(x)\biggl]\,,
\end{eqnarray}
where "ext" and "pnt" refer to the extended-nucleus and the point-nucleus model, respectively.
The partial-wave expansion of the first part converges very slowly, so we used the approximate
expression for the point-nucleus effective potential from Eq.~(\ref{eq:4}) to evaluate this term.
The second term was calculated directly according to Eq.~(\ref{eq:33}), by taking the difference
of the extended-nucleus and point-nucleus Dirac-Coulomb Green function. In this case, the
partial-wave expansion converges rapidly; it was sufficient to take into account just the three first
terms of the expansion.

\section{Results and discussion}
\label{sec:3}

\begin{figure}
\centerline{
\resizebox{0.8\textwidth}{!}{%
  \includegraphics{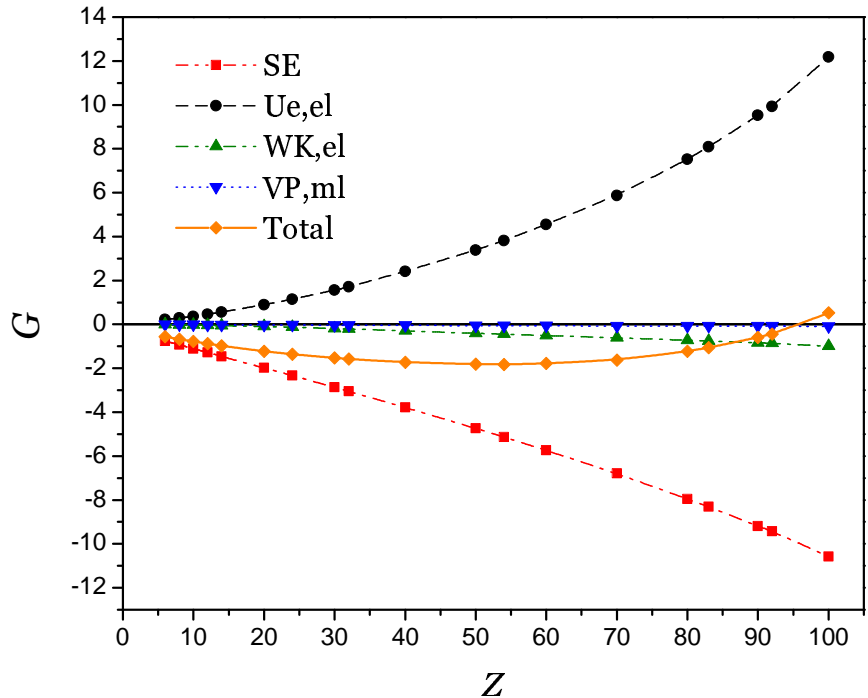}
}}
 \caption{Nuclear-size QED corrections to the $1s$ bound-electron $g$ factor. Notations are the same as in
 Table~\ref{tab:qed_fns}.
 \label{fig:qed_fns}}
\end{figure}

Numerical results of our calculations of the nuclear-size QED corrections to the $1s$
bound-electron $g$ factor of hydrogen-like ions are presented in Table~\ref{tab:qed_fns}
and plotted in Fig.~\ref{fig:qed_fns}. Table~\ref{tab:qed_fns} also presents comparison with the
results obtained in the previous studies \cite{beier:00:rep,yerokhin:04}. Our results allow us also to
check the accuracy of the approximate relation (\ref{eq:33a}) between the nuclear-size vacuum-polarization
correction to the $g$-factor and the binding energy. Our conclusion is that this
relation yields a rather crude approximation. It holds with accuracy of about 5\% for $Z \ge 80$
and 10\% for $Z\ge 40$.

We observe that the dominant contribution to the nuclear-size QED correction comes from the
self-energy and the Uehling part of the vacuum-polarization. These two contributions are of
different sign and largely cancel each other. In the low-$Z$ region, the self-energy
dominates over the vacuum-polarization, but in the high-$Z$ region both corrections have
the same order of magnitude. The resulting nuclear-size QED correction turns out to be rather small
in the whole region of the nuclear charges.

\begin{table*}
\caption{Nuclear-size QED corrections to the $1s$ bound-electron $g$ factor, expressed in terms of
 the function $G_{\rm NQED}(Z)$ defined by Eq.~(\ref{eq:1}). Abbreviations are as follows: "SE" denotes
self-energy contribution, "Ue,el" denotes the Uehling electric-loop vacuum-polarization correction, "WK,el" stands for
the Wichmann-Kroll electric-loop vacuum-polarization correction, and "VP,ml" denotes the magnetic-loop
vacuum-polarization contribution.
 \label{tab:qed_fns} }
\begin{center}
\begin{tabular}{lc.....}
\hline\\[-0.5cm]
                $Z$ & R [fm]
                & \multicolumn{1}{c}{SE}
                & \multicolumn{1}{c}{Ue,el}
                & \multicolumn{1}{c}{WK,el}
                & \multicolumn{1}{c}{VP,ml}
                                         & \multicolumn{1}{c}{Total} \\
\hline\\[-9pt]
  6 & 2.4703 & -0.7x60\,(5)     &   0.1x80       & -0.0x11    & -0.0x1\,(1)      & -0.6x0\,(1)      \\
  8 & 2.7013 & -0.9x30\,(4)     &   0.2x57       & -0.0x19    & -0.0x1\,(1)      & -0.7x0\,(1)      \\
 10 & 3.0053 & -1.1x05\,(3)     &   0.3x40       & -0.0x28    & -0.0x14\,(9)     & -0.8x07\,(9)     \\
 12 & 3.0568 & -1.2x80\,(2)     &   0.4x33       & -0.0x40    & -0.0x18\,(8)     & -0.9x05\,(8)     \\
 14 & 3.1223 & -1.4x58\,(2)     &   0.5x35       & -0.0x53    & -0.0x19\,(5)     & -0.9x96\,(5)     \\
 20 & 3.4764 & -1.9x84\,(2)     &   0.8x72       & -0.0x99    & -0.0x27\,(2)     & -1.2x37\,(3)     \\
    &        & -1.9x8\,^a \\
 24 & 3.6424 & -2.3x38\,(2)     &   1.1x31       & -0.1x34    & -0.0x32\,(1)     & -1.3x72\,(2)     \\
 30 & 3.9286 & -2.8x72\,(2)     &   1.5x60       & -0.1x92    & -0.0x38\,(1)     & -1.5x42\,(2)     \\
 32 & 4.0744 & -3.0x50\,(1)     &   1.7x08       & -0.2x11    & -0.0x40\,(1)     & -1.5x93\,(1)    \\
 40 & 4.2696 & -3.7x87\,(1)     &   2.4x00       & -0.2x98    & -0.0x49\,(1)     & -1.7x33\,(1)    \\
 50 & 4.6543 & -4.7x36\,(1)     &   3.3x77       & -0.4x05    & -0.0x57\,(1)     & -1.8x21\,(1)    \\
    &        & -4.7x5\,^a \\
 54 & 4.7866 & -5.1x30\,(1)     &   3.8x15       & -0.4x49    & -0.0x60\,(1)     & -1.8x23\,(1)    \\
 60 & 4.9118 & -5.7x43\,(1)     &   4.5x44       & -0.5x16    & -0.0x65\,(1)     & -1.7x80\,(1)    \\
 70 & 5.3115 & -6.7x94\,(1)     &   5.8x60       & -0.6x16    & -0.0x71\,(1)     & -1.6x21\,(1)    \\
 80 & 5.4633 & -7.9x51\,(1)     &   7.5x27       & -0.7x30\,(1) & -0.0x77\,(1)     & -1.2x32\,(1)    \\
 83 & 5.5211 & -8.3x15\,(1)     &   8.0x90       & -0.7x65\,(1) & -0.0x78\,(1)     & -1.0x68\,(1)    \\
 90 & 5.7100 & -9.1x89\,(1)     &   9.5x28       & -0.8x47\,(1) & -0.0x82\,(2)     & -0.5x90\,(2)    \\
    &        & -9.1x86\,^b        &   9.4x94\,^b     & -0.8x43\,^b    & -0.0x62\,^b \\
    &        & -9.1x7\,^a     \\
 92 & 5.8569 & -9.4x27\,(1)     &   9.9x27       & -0.8x66\,(1) & -0.0x83\,(2)     & -0.4x49\,(2)    \\
100 & 5.8570 & -10.5x78\,(1)    &  12.1x73       & -0.9x92\,(2) & -0.0x86\,(2)     &  0.5x18\,(3)     \\
\hline
\end{tabular}
\end{center}
$^a$ Ref.~\cite{yerokhin:04}, shell nuclear model;
$^b$ Ref.~\cite{beier:00:rep}, $R = 5.802$.
\end{table*}

We now turn to the experimental consequences of our calculations. Table~\ref{tab:is} presents
theoretical results for the nuclear-dependent part of the $1s$ bound-electron $g$ factor
and for the isotope shift of the bound-electron $g$ factor for several hydrogen-like ions.
The leading-order nuclear-size contribution (labeled as "N") and the nuclear-size self-energy ("NSE") and
vacuum-polarization ("NVP") corrections are taken from
Tables~\ref{tab:fns} and \ref{tab:qed_fns}. The uncertainty
of the leading nuclear-size correction represents the model dependence of the calculation, defined
as the difference of the results obtained with the Fermi and the homogeneously charged sphere models.

The data presented in Table~\ref{tab:is} for the recoil corrections were taken from the previous studies.
The recoil correction of first order in the electron-to-nucleus mass ratio $m/M$ (labeled as "REC")
was calculated to all orders in $\Za$ in Ref.~\cite{shabaev:02:recprl}. The radiative and higher-order recoil
corrections are known to the leading order in $\Za$ \cite{eides:97:apn}
\begin{eqnarray}
& \Delta g_{\rm REC,QED}  = -\frac{\alpha}{\pi}\,\frac{m}{M}\,\frac{(\Za)^2}{3}\,,
\\
& \Delta g_{\rm REC2} = -\left( \frac{m}{M} \right)^2\,(\Za)^2\,(1+Z)\,.
\end{eqnarray}

Apart of the nuclear-size and nuclear-recoil effects, the bound-electron $g$ factor is also influenced by
various nuclear-structure effects. Out of those, the nuclear polarization is probably the largest.
The correction to the bound-electron $g$ factor due to the nuclear polarization was calculated for several
ions in Ref.~\cite{nefiodov:02:prl}. Unfortunately, the data presented in that work are not sufficient
for our compilation in Table~\ref{tab:is}. Because of this, we approximate the uncertainty due to
the nuclear-polarization effect as 50\% of the uncertainty
due to the model-dependence of the nuclear-size effect. We observed that for most cases calculated
in Ref.~\cite{nefiodov:02:prl}, the nuclear polarization correction is (crudely) consistent with this simple
estimate. In particular, for $^{208}_{82}$Pb, our estimate yields $4\times 10^{-7}$, whereas the
numerical results of Ref.~\cite{nefiodov:02:prl} is $2.2\times 10^{-7}$; for $^{84}_{36}$Kr, our estimate yields
$1\times 10^{-9}$, to be compared with $1.2\times 10^{-9}$ of Ref.~\cite{nefiodov:02:prl}.

The final results presented in Table~\ref{tab:is} for the nuclear-dependent part of the bound-electron $g$ factor and for the
isotope shift  have two uncertainties. The first one is the
estimation of the model dependence of the nuclear-size correction, whereas the second one is the
estimate of the nuclear polarization effect. The
errors due to the experimental values of the nuclear radii are not shown explicitly in
the table, but they can be easily deduced from the $R$ dependence of the results, see
Eq.~(\ref{eq:04}).

We note that the uncertainty of the model dependence of the nuclear-size contribution diminishes significantly in
the isotope-shift difference, but not that of the nuclear polarization. The nuclear
polarization effect can vary significantly between the isotopes, so one cannot expect a high degree
of cancelation in this case. The error due to nuclear polarization dominates in the
theoretical isotope shift and currently
sets the limit to possible determinations of the difference of the charge
radii from the bound-electron $g$ factor measurements.

Summarizing, in the present investigation we calculated the
finite nuclear-size effect on the leading bound-electron $g$ factor and on the one-loop QED
corrections to the bound-electron $g$ factor in hydrogen-like atoms.
The calculation was performed to all orders
in the nuclear binding strength parameter $\Za$ and for the Fermi model of the nuclear charge distribution.
Combined with the previous calculations of the nuclear recoil effect, our investigation
yields theoretical values for the isotope shift of the $1s$
bound-electron $g$ factor that can be used for determination of the isotope differences of the nuclear charge radii from measurements
of the bound-electron $g$ factor in hydrogen-like ions.

\section*{Acknowledgement}

The work presented in the paper was supported by the Alliance Program of the Helmholtz Association
(HA216/EMMI). Z.H. acknowledges insightful conversations with Jacek Zatorski and Natalia Oreshkina.

\section*{References}


\newpage

\begin{table*}
\caption{Nuclear-dependent contributions and the isotope shifts (IS)
of the $1s$ bound-electron $g$ factor for several hydrogen-like ions,
multiplied by $10^6$.
 \label{tab:is} }
\begin{center}
\vspace*{-0.5cm}
\begin{tabular}{l...}
\hline\\[-0.5cm]
      &  \multicolumn{1}{c}{$^{40}{\rm Ca}^{19+}$} & \multicolumn{1}{c}{$^{44}{\rm Ca}^{19+}$} & \multicolumn{1}{c}{{\rm IS}} \\
      \hline
$R$   &    3.476x4      &    3.515x5 \\
$m/M\times 10^5$ &  1.373\,x11   &      1.248\,x35 \\
N     &    0.113\,x029\,(52) &  0.115\,x556\,(52) &   0.002\,x527\,(1) \\
NSE   &   -0.000\,x521     & -0.000\,x533     &  -0.000\,x012 \\
NVP   &    0.000\,x196     &  0.000\,x200     &   0.000\,x004 \\
REC   &    0.297\,x378     &  0.270\,x358     &  -0.027\,x020 \\
REC,QED  &-0.000\,x226     & -0.000\,x206     &   0.000\,x021 \\
REC2  &   -0.000\,x084     & -0.000\,x070     &   0.000\,x015 \\
Total &    0.409\,x772\,(52)(26) &   0.385\,x306\,(52)(26) &  -0.024\,x466\,(1)(26)\\
\hline \hline
      &  \multicolumn{1}{c}{$^{86}{\rm Kr}^{35+}$} & \multicolumn{1}{c}{$^{78}{\rm Kr}^{35+}$} & \multicolumn{1}{c}{{\rm IS}} \\
      \hline
$R$   &    4.18x36       &   4.20x32 \\
$m/M\times 10^5$ &  0.63x87    &     0.70x42 \\
N     &    2.25x62\,(20)   &   2.27x66\,(20)  &    0.02x0\,39\,(2) \\
NSE   &   -0.01x79       &  -0.01x81      &   -0.00x0\,16 \\
NVP   &    0.00x91       &   0.00x92      &    0.00x0\,08 \\
REC   &    0.47x31       &   0.52x17      &    0.04x8\,53 \\
REC,QED&  -0.00x03       &  -0.00x04      &   -0.00x0\,04 \\
REC2  &   -0.00x01       &  -0.00x01      &   -0.00x0\,02 \\
Total &    2.72x01\,(20)(10)   &   2.78x89\,(20)(10)  &    0.06x8\,79\,(2)(100) \\
\hline \hline
      &  \multicolumn{1}{c}{$^{128}{\rm Xe}^{53+}$} & \multicolumn{1}{c}{$^{136}{\rm Xe}^{53+}$} & \multicolumn{1}{c}{{\rm IS}} \\
      \hline
$R$    &     4.77x55     &     4.79x91 \\
$m/M\times 10^5$ &  0.42x90   &      0.40x37 \\
N      &  23.38x5\,(32) &   23.59x7\,(32) &   0.21x1\,79\,(45) \\
NSE    &  -0.27x7     &   -0.28x1     &  -0.00x2\,52 \\
NVP    &   0.18x0     &    0.18x1     &   0.00x1\,63 \\
REC    &   0.80x8     &    0.76x1     &  -0.04x7\,62 \\
REC,QED&  -0.00x1     &    0.00x0     &   0.00x0\,03 \\
REC2   &   0.00x0     &    0.00x0     &   0.00x0\,02 \\
Total  &  24.09x4\,(32)(16) &   24.25x7\,(32)(16) &   0.16x3\,3\,(5)(160)\\
\hline \hline
      &  \multicolumn{1}{c}{$^{204}{\rm Pb}^{81+}$} & \multicolumn{1}{c}{$^{208}{\rm Pb}^{81+}$} & \multicolumn{1}{c}{{\rm IS}} \\
      \hline
$R$   &      5.4x794    &      5.5x010 \\
$m/M\times 10^5$ &  0.2x690 &         0.2x638 \\
N      &   450.0x8\,(88) &    452.8x9\,(88) &   2.8x08\,(11) \\
NSE    &    -8.5x7     &     -8.6x2     &  -0.0x53\\
NVP    &     7.3x9     &      7.4x3     &   0.0x46\\
REC    &     1.7x6     &      1.7x2     &  -0.0x34\\
REC,QED&     0.0x0     &      0.0x0     &   0.0x00\\
REC2  &      0.0x0     &      0.0x0     &   0.0x00\\
Total &    450.6x6\,(88)(44) &    453.4x2\,(88)(44) &   2.7x67\,(11)(440)\\
\hline
\end{tabular}
\end{center}
\end{table*}

\end{document}